\documentclass[journal]{IEEEtran}
\usepackage[dvips]{graphicx}
\begin{document}

\title{The ATLAS Forward Physics Project}
\author{Christophe Royon
\thanks{DSM/IRFU/SPP, CEA Saclay, France;
On behalf of the AFP collaboration}}

\maketitle

\begin{abstract}
We describe the main components of the ATLAS Forward Physics project, namely the
movable beam pipe, the tracking and timing detectors which allow to detect
intact protons in the final state at the LHC. The position detector is composed on 6 layers
of 3D silicon detectors readout by FE-I4 chips developped for ATLAS. The fast timing
detector is built from a quartz-based Cerenkov detector coupled to a
microchannel plate photomultiplier tube, followed by the 
electronic elements  that amplify, measure, and record the time of the event 
along with a stabilized reference clock signal, ensuring a time resolution of
10-15 picoseconds.
\end{abstract}

We describe the proposal to install the ATLAS Forward Proton (AFP)
detector in order to detect intact  protons at 206 and 214 meters on both side of
the ATLAS experiment~\cite{LOI}. 
This one arm will consist of two sections (AFP1 and AFP2) contained in a special design of
beampipe described in Section I. In the first section (AFP1), in one pocket of the beampipe,
a tracking station composed by 6 layers of Silicon detectors described in
Section II will be deployed.
The second station AFP2 will contains 
another tracking station similar to the one already described and a timing detector described in 
Section III.
The aim of this setup, mirrored by an identical arm placed on the opposite side of the ATLAS IP,
will be to tag the protons emerging intact from the $pp$ interactions so allowing ATLAS to exploit
the program of diffractive and photoproduction processes~\cite{chr}. This device is sensitive to diffractive
masses in the ATLAS detectors between 350 and 1.4 TeV. The advantage of this apparatus from the
physics point of view is that the system is fully constrained: we measure all the particles in the
final state, namely the two intact protons and the particle produced ($W$, $Z$, the Higgs boson or the
SUSY particles). The mass and the kinematical properties of the produced particles can be computed
using the information on the tagged protons.

\section{Movable beam pipes}
\begin{figure}
\centering
\includegraphics[width=4.in]{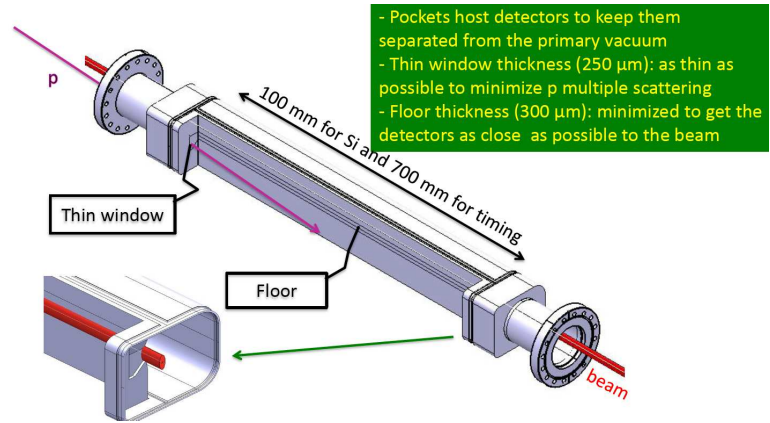}
\caption{Scheme of the movable beam pipe.}\label{Figa}
\end{figure}

The idea of movable Hamburg beam pipes is quite simple~\cite{HBP}: a larger section of the 
LHC beam pipe than the usual one
can move close to the beam using bellows so that the detectors located at its edge (called pocket)
can move close
to the beam by about 2.5 cm when the beam is stable (during injection, the detectors are in
parking position). 
In its design, the 
predominant aspect is the minimization of the thickness of the portions 
called floor and window (see Fig.~ \ref{Figa}). Minimizing the depth of the floor ensures that the
detector can go as close to the beam as possible allowing us to detect protons scattered at very small
angles, while minimizing the depth of the thin window is important to keep the protons intact and
to reduce the impact of multiple interactions. 
Two configurations exist for the movable beam pipes: the first one at 206 m form the ATLAS interaction
point hosts a Si detector 
(floor length of about 100 mm) and the second one (floor 
length of about 700 mm) the timing and the Si detectors.

\section{3D Silicon detectors}
The purpose of the tracker system is to measure points along the trajectory of 
beam protons that are deflected at small angles as a result of collisions. 
The tracker when combined with the LHC dipole and quadrupole magnets,
forms a powerful momentum spectrometer. Silicon tracker stations will be
installed in  Hamburg beam pipes (HBP) at $\pm$ 206 and $\pm$ 214 m from the 
ATLAS. 

The key requirements for the silicon  tracking system at 220~m are:
\begin{itemize}
\item Spatial resolution of $\sim$ 10 (30) $\mu$m per detector station in $x$ ($y$)
\item Angular resolution for a pair of detectors of about 1~$\mu$rad
\item High efficiency over the area of 20~mm $\times$ 20~mm corresponding to the distribution of
diffracted protons
\item Minimal dead space at the edge of the sensors allowing us to measure the scattered protons at
low angles
\item Sufficient radiation hardness in order to sustain the radiation at high luminosity
\item Capable of robust and reliable operation at high LHC luminosity 
\end{itemize}

The basic building unit of the AFP detection system is a module consisting of 
an assembly of a sensor array, on-sensor read-out chip(s), electrical services,
data acquisition and detector control system. The module will be
mounted on the mechanical support with embedded cooling and other necessary
services. The sensors are double sided 3D 50$\times$250 micron pixel detectors with slim-edge 
dicing built by FBK and CNM. The sensor efficiency has been measured to be close to 100\% 
over the full size in beam tests. A possible upgrade of this device will be to use 3D edgless Silicon
detectors built in a collaboration between SLAC, Manchester, Oslo, Bergen...
A new 
front-end chip FE-I4 has been developed for the Si detector by the Insertable B Layer (IBL)
collaboration in ATLAS. The FE-I4 integrated circuit contains 
readout circuitry for 26 880 hybrid pixels arranged in 80~columns on 250~$\mu$m
pitch by 336 rows on 50 $\mu$m pitch, and covers an area of about 19 mm 
$\times$ 20 mm. It is designed in a 130 nm feature size bulk CMOS process. 
Sensors must be DC coupled to FE-I4 with negative charge collection. The FE-I4 
is very well suited to the AFP requirements: the granularity of cells provides 
a sufficient spatial resolution, the chip is radiation hard enough 
(up to a dose of 3~MGy), 
and the size of the chip is sufficiently 
large that one module can be served by just  one chip. 

The dimensions of the individual cells in the FE-I4 chip are 50 $\mu$m $\times$
250 $\mu$m in the  $x$ and $y$ directions, respectively.
Therefore to achieve the required position resolution in the
$x$-direction of $\sim$ 10 $\mu$m, six layers with sensors are required
(this gives  50/$\sqrt{12}$/$\sqrt{5}\sim 7$ $\mu$m in $x$ and roughly 5 times 
worse in $y$). Offsetting planes alternately to the left and right by one half 
pixel will give a further reduction in resolution of at least 30\%. 
The AFP sensors are expected to be exposed to a dose of 30~kGy
per year at the full LHC luminosity of 10$^{34}$cm$^{-2}$s$^{-1}$.

\section{Timing detectors}

A fast timing system that can precisely measure the time difference between 
outgoing scattered protons is a key component of the AFP detector.  The 
time difference is equivalent to a constraint on the event vertex, thus 
the AFP timing detector can be used to reject overlap background by 
establishing that the two scattered protons did not originate from the same 
vertex as the the central system.  The final timing 
system should have the following characteristics~\cite{timing}: 
\begin{itemize}
\item 10 ps or better resolution (which leads to a factor 40 rejection on pile up 
background)
\item Efficiency close to 100\% over the full detector coverage
\item High rate capability (there is a bunch crossing every 25 ns at the nominal LHC)
\item Enough segmentation for multi-protin timing
\item Level trigger capability
\end{itemize}

\begin{figure}
\centering
\includegraphics[width=3.5in]{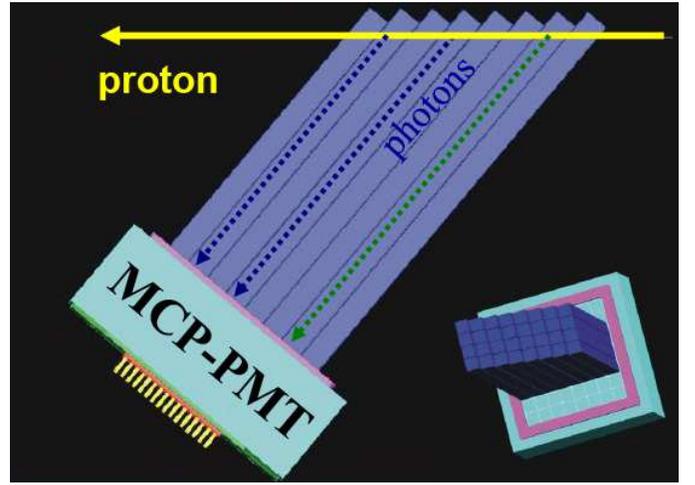}
\vspace*{-0.1in}
  \caption{A schematic diagram of the QUARTIC fast timing detector.
   } \label{fig5_timesys}
\end{figure}

\begin{figure}
\centering
\includegraphics[width=3.5in]{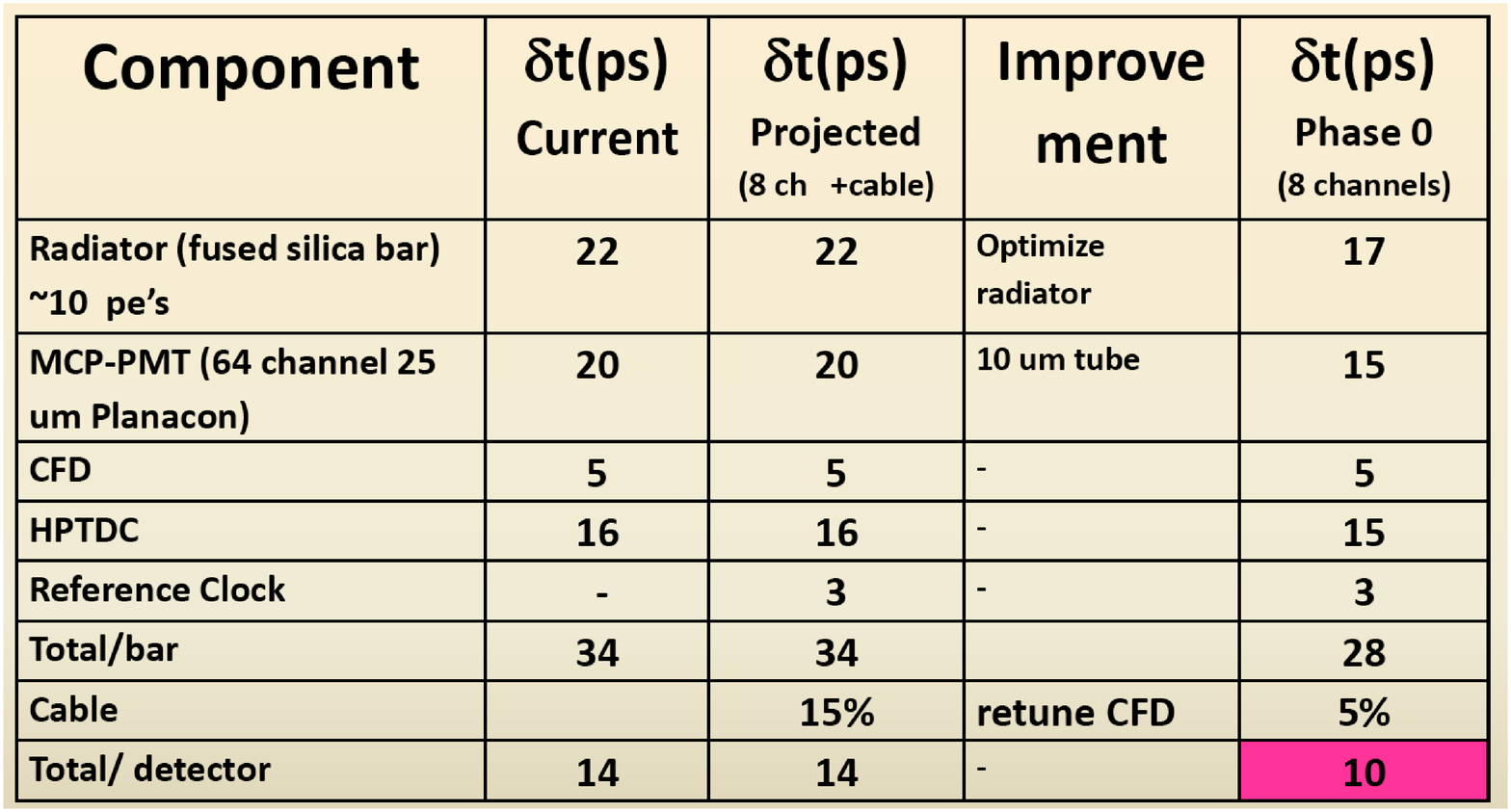}
\vspace*{-0.1in}
  \caption{Different components of the timing detector resolution.
   } \label{timresol}
\end{figure}

Figure~\ref{fig5_timesys} shows a schematic overview of the first proposed 
timing system, consisting of a quartz-based Cerenkov detector coupled to 
a microchannel plate photomultiplier tube (MCP-PMT), followed by the 
electronic elements  that amplify, measure, and record the time of the event 
along with a stabilized reference clock signal. The
QUARTIC detector consists of  an array of 8$\times$4 fused
silica bars ranging in length from about 8 to 12 cm and oriented at the
average Cerenkov angle.  A proton that is sufficiently deflected from the beam axis will pass 
through a row of eight bars emitting Cerenkov photons providing an overall time resolution that is 
approximately $\sqrt{8}$ times smaller than the single bar resolution of about 30 ps, 
thus approaching the 10 ps resolution goal. Prototype tests have generally been performed on one row 
(8 channels) of 5 mm $\times$ 5 mm pixels, while the initial detector is foreseen to have four rows
to obtain full acceptance out to 20  mm from the beam. The beam tests lead to a time resolution per
bar of the order of 34 ps. The different componenets of the timing resolution are given in
Fig.~\ref{timresol}. The upgraded design of the timing detector has equal rate pixels, 
and we plan to reduce the the width of detector bins
close to the beam, where the proton density is highest. 

At higher luminosity of the LHC (phase I starting in 2019), higher pixelisation of the timing detector
will be required. For this sake, a R\&D phase concerning timing detector developments based on SiPMs,
APDs, quartz fibers, diamonds has been started in Saclay. In parallel, a new timing readout chip has
been developed. It uses waveform sampling methods which give the best possible timing resolution. The
aim of this chip called SAMPIC~\cite{sampic} is to obtain sub 10 p timing resolution, 1GHz input bandwidth, no dead
time at the LHC, and data taking at 2 Gigasamples per second. The cost per channel is estimated to be
of the order for 10\$ which a considerable improvement to the present cost of 10,000 \$ per channel,
allowing us to use this chip in medical applications such as PET imaging detectors. The holy grail of
imaging 10 picosecond PET detector seems now to be feasible: with a resolution better than 20 ps,
image reconstruction is no longer necessary and real-time image formation becomes
possible~\cite{manjit}.

\end{document}